\documentclass[a4paper]{article}
\usepackage{INTERSPEECH2021}

\usepackage{multirow,hhline} 

\title{Evaluation of Speaker Anonymization on Emotional Speech}

\name{Hubert Nourtel$^{1 \star}$\quad
Pierre Champion$^{1,2 \star}$\quad
Denis Jouvet$^{1}$\quad
Anthony Larcher$^{2}$\quad
Marie Tahon$^{2}$}

\address{
$^{1}$ Université de Lorraine, CNRS, Inria, LORIA, {\normalsize F-54000 Nancy, France}\\
$^{2}$ LIUM - EA4023, Le Mans Université, {\normalsize Avenue Olivier Messiaen,72085 LE MANS CEDEX 9, France} \\
$^{\star}$ equal contribution from authors
}
\email{hubert.nourtel@inria.fr}

\begin{document}

\maketitle
\begin{abstract}
Speech data carries a range of personal information, such as the speaker's identity and emotional state.
These attributes can be used for malicious purposes.
With the development of virtual assistants, a new generation of privacy threats has emerged.
Current studies have addressed the topic of preserving speech privacy.
One of them, the VoicePrivacy initiative aims to promote the development of privacy preservation tools for speech technology.
The task selected for the VoicePrivacy 2020 Challenge (VPC) is about speaker anonymization. The goal is to hide the source speaker's identity while preserving the linguistic information. The baseline of the VPC makes use of a voice conversion.
This paper studies the impact of the speaker anonymization baseline system of the VPC on emotional information present in speech utterances.
Evaluation is performed following the VPC rules regarding the attackers' knowledge about the anonymization system.
Our results show that the VPC baseline system does not suppress speakers' emotions against informed attackers. When comparing anonymized speech to original speech, the emotion recognition performance is degraded by 15\% relative to IEMOCAP data, similar to the degradation observed for automatic speech recognition used to evaluate the preservation of the linguistic information.

\end{abstract}

\section{Introduction}
Voice-controlled applications, such as smart speakers, have become widely popular.
Large amount of data is required to train such applications. This motivates service providers to collect, process, and store personal data in centralized servers.
Voice is one of the most sensitive modalities as it encapsulates many discernible attributes of a speaker such as age, gender, health, personality traits, socioeconomic status, geographical origin, biometric identity, moods, and emotions \cite{speech_characterisation, privacy_implication_voice}.
Given that speech data falls under the category of personal data \cite{nautschGDPRSpeechData2019}, speech privacy-preserving solutions are becoming increasingly important.
Additionally, recent regulations, e.g., the General Data Protection Regulation (GDPR)~\cite{gdpr} in the European Union, emphasize on privacy preservation and protection of personal data.
The research reported in this article has been done using the Voice privacy Challenge (VPC) framework \cite{tomashenkoVoicePrivacy2020Challenge} which is one of the first attempts of the speech community to evaluate research on this topic by producing dedicated protocols, metrics, datasets, and baselines.

The goal of the VPC system is to anonymize the speaker. This task is performed to suppress the personally identifiable paralinguistic information from a speaker's speech utterance while maintaining the linguistic content.
The VPC baseline system uses a speaker anonymization approach \cite{fangSpeakerAnonymizationUsing2019} based on x-vectors and voice conversion.
The quality of anonymization in the VPC is assessed using a speaker verification system, which evaluates the speaker concealing capability (privacy metric) and using an automatic speech recognition system to evaluate the preservation and intelligibility of the linguistic content (utility metric) \cite{tomashenkoVoicePrivacy2020Challenge}.
In this work, we investigate the extent to which an utterance's emotional content can be retrieved after anonymization.

Speaker recognition and voice privacy usually focus on so-called "neutral" speech. However, in spontaneous expressive speech, the audio signal carries speaker information, linguistic content and emotional cues.
The anonymization process can be altered by emotional speech, for which the speech signal strongly differs from the "neutral" speech.
Human emotion is usually described in psychological theories using diverse and complementary theories~\cite{Russell1980,Scherer2005}.
For a long time, the collection of emotional data mainly focused on acted and semantically controlled data from a few speakers~\cite{emo-db}.
However, the actual trend is to capture the diversity of humans expressively in real-life conditions in order to model social aspects or induced interactions such as laughter~\cite{Devillers2015} or disfluencies~\cite{Gilmartin2016}.
The Interactive Emotional Dyadic Motion Capture (IEMOCAP) dataset~\cite{Busso08} is in-between acted and spontaneous speech and has the advantage of being used as a benchmark in the community.

Linguistic cues mainly rely on the words pronounced by the speaker, while paralinguistic cues are directly related to the acoustic content of the speech signal.
More precisely, prosodic features such as the fundamental frequency (F0), intensity and rhythm, are often considered as the most important cues in the field of speech emotion recognition (SER).
Although most SER systems intend to capture prosody in input, another option is to extract Mel frequency cepstral coefficients (MFCCs)~\cite{macary2020multicorpus} or even spectrograms~\cite{Li2019discriminative}.
The HUMAINE association also took an inventory of acoustic features in the CEICES initiative~\cite{batiner_2006_ceices} which conducts to a set of a hundred descriptors selected over several corpora with various techniques~\cite{GEMAPS}.
These features have the advantage of being easily interpretable. However, their extraction in degraded signals is error-prone.
The use of input pre-trained features, i.e., embeddings extracted with neural models trained for speech processing tasks different from SER, are currently extensively used.
The advantage of such an approach is to benefit from a large amount of data from a different task such as automatic speech recognition (ASR) \cite{Yeh2020asr} or speaker recognition~\cite{Pappagari2020XVectorsME}.

Remarkably few works have handled the problem of privacy preservation in the context of emotional speech.
In \cite{dias2018emotion}, the authors proposed distance-preserving hashing techniques and homomorphic encryption to protect sensitive data such as emotions.
Generative adversarial networks have been used as an intermediate layer between users and cloud services to sanitize the input speech~\cite{aloufi2019emotionless}.
We aim to investigate how applying an anonymization process on emotional data, which is supposed to hide the speaker identity, impacts SER performance.



The paper is organized as follows.
Section \ref{sec:voice_privacy} presents the anonymization framework. It recalls the VPC baseline system, details F0 transformation enhancements, and presents the attack scenarios.
Section \ref{sec:experiments} details the experiments conducted and the evaluation protocol with respect to the emotions and discusses the results.
A conclusion ends the paper.

\section{Anonymization framework}
\label{sec:voice_privacy}
This section presents the speaker anonymization baseline system of the Voice Privacy Challenge and the attack scenarios.

\subsection{The VPC speaker anonymization system}

\begin{figure}[ht]
  \centering
  \includegraphics[width=1.00\linewidth]{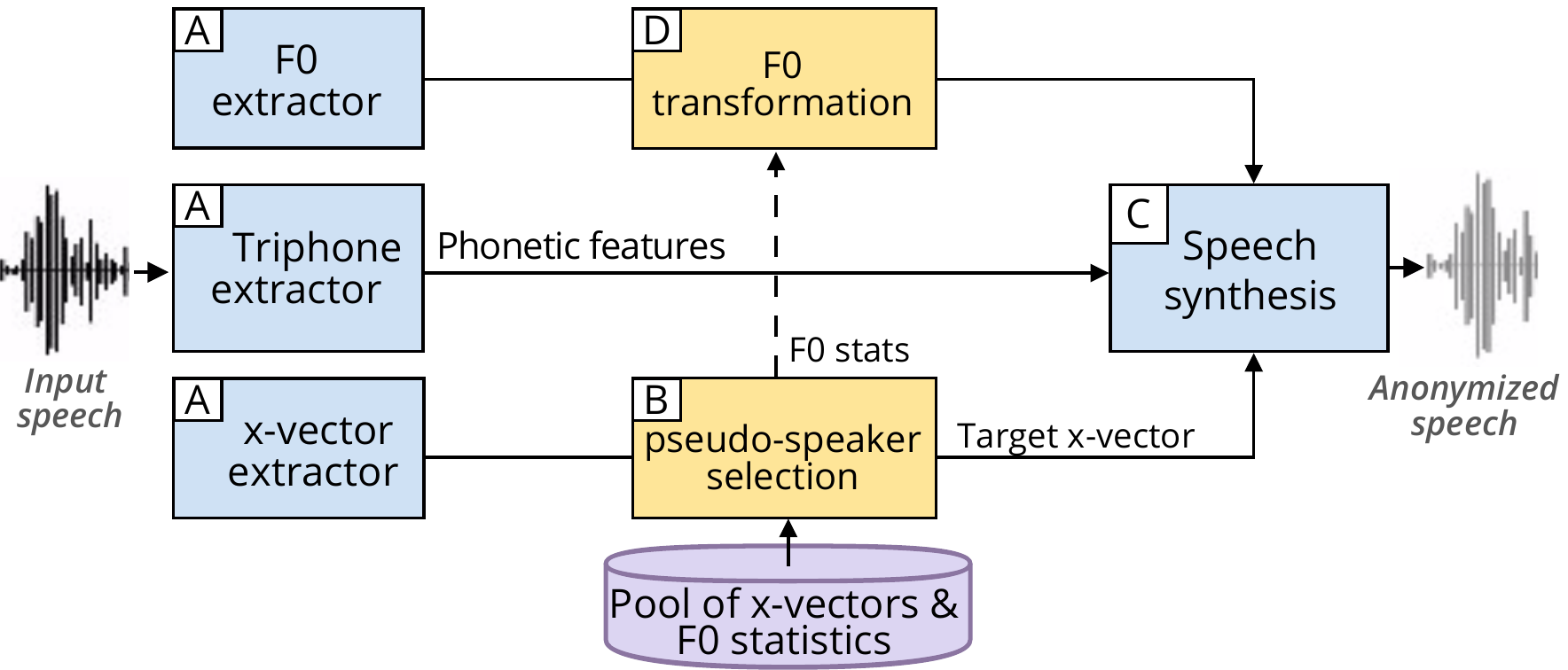}
  \caption{The Voice Privacy speaker anonymization pipeline. Modules A, B, and C are parts of the baseline model. Module D is an enhancement used to transform the F0 values.
  }
  \label{fig:baseline.png}
  \vspace{-0.3cm}
\end{figure}

The baseline system introduced in \cite{fangSpeakerAnonymizationUsing2019} aims at separating speaker identity and linguistic content from an input speech utterance.
Assuming that those features are disentangled, an anonymized speech waveform is generated by altering only the features that encode the speaker's identity.
The anonymization system depicted in Figure~ \ref{fig:baseline.png} can be decomposed into three groups of modules.
Modules from the \textit{group A} extract different features from the source signal: the fundamental frequency (F0), the phonetic features encoding articulation of speech sounds and the speaker's x-vector.
\textit{The module B} derives a new pseudo-speaker target x-vector identity. The x-vector from each source input speaker is compared to a pool of external x-vectors to select the 200 furthest x-vectors; 100 of them are randomly selected and averaged to create an anonymized pseudo-speaker x-vector identity.
Finally, \textit{the module C} synthesizes a speech waveform from the target x-vector together with the original phonetic features and F0.
Speaker anonymization is achieved through the selection of the pseudo-speaker target x-vector.
The triphone extractor has been trained on the \textit{train-clean-100} and \textit{train-other-500} subsets of LibriSpeech. The x-vector extractor has been trained on VoxCeleb-1,2. The speech synthesis system has been trained on the \textit{train-clean-100} subset of LibriTTS. Finally, the \textit{train-other-500} subset of LibriTTS has been used to create a pool of x-vector and F0 statistics.

\subsection{The F0 transformations}

In the original VPC anonymization system, the F0 values extracted from the source speech are directly used (unchanged) by the speech synthesizer, even though a different pseudo-speaker target x-vector is selected.
Multiple studies have investigated F0 conditioned voice conversion \cite{F0_Bahmaninezhad2018ConvolutionalNN,f0_Huang2020UnsupervisedRD,F0_Qian2020F0ConsistentMN,F0_IndividualityPreservingSM,pitch_contour_modeling}.
They conclude that modifying the F0 improves the quality of the converted voice.
Motivated by those results, and also by the fact that emotions undoubtedly affect intonation, we propose to modify the F0 values of a source utterance from a given speaker (cf. module D in Figure~\ref{fig:baseline.png}) by using a linear transformation and a random warping.

\subsubsection{F0 linear transformation}
In this method \cite{F0_Bahmaninezhad2018ConvolutionalNN,f0_Huang2020UnsupervisedRD}, to feed the synthesizer with F0 values close to the pseudo-speaker selected, the F0 features of the source speaker are transformed using a linear transformation:
\begin{equation}
    \log \hat{x}_{t}=\mu_{y}+\frac{\sigma_{y}}{\sigma_{x}}\left(\log x_{t}-\mu_{x}\right)
\end{equation}
where $x_{t}$ represents the F0 of the source speaker at frame $t$, $\mu_{x}$ and $\sigma_{x}$ represent the mean and standard deviation of the log-scaled F0 for the source speaker computed on all his/her utterances. $\mu_{y}$ and $\sigma_{y}$ represent the mean and standard deviation of the log-scaled F0 for the pseudo-speaker.
The linear transformation and statistical calculation are only performed on voiced frames.
The mean and standard deviation for the target pseudo-speaker is calculated by averaging the F0 of voiced frames from the 100 speakers selected to derive the pseudo-speaker x-vector.

\subsubsection{F0 random warping}

\label{subsubsec:f0_random}
In this method \cite{pitch_contour_modeling,frequency_range}, the contour of the F0 values are randomly modified to increase or decrease the range of the F0 variation using a warping factor:
\begin{equation}
    \hat{x}_{t}=\mu_{x}+\left(x_{t} - \mu_{x}\right) \times \alpha
\end{equation}
where $\alpha$ is sampled from a uniform distribution between $0.8$ and $1.2$, $x_{t}$ represents the F0 value at frame $t$ and $\mu_{x}$ represents the mean F0 of the utterance to transform. The $\alpha$ warping factor is randomly sampled for each utterance. This F0 random warping is applied after the F0 linear transformation.

\subsection{The VPC attack scenarios}
\label{subsec:VPC_scenars}

In the Voice Privacy Challenge, multiple sets of tests were performed depending on the attacker's knowledge of the anonymization algorithm.
In this work, we focus on the \textit{Ignorant}, and the \textit{Informed} attacker scenarios  \cite{EvaluatingVoiceConversionbased2019,brij_journal}.
In the \textit{Ignorant} scenario, the attacker is unaware that speech is transformed.
Thus, privacy measurement is assessed using models trained on original, non-anonymized data, while the evaluation is performed using anonymized data.
This mismatch leads to the measurement of a rather good anonymization performance.
On the opposite, the \textit{Informed} attacker scenario is entirely aware of the anonymization algorithm.
Such attackers are able to anonymize a training dataset in the same manner as the service provider.
This anonymized dataset is later used to train the evaluation model.

\section{Experiments}
\label{sec:experiments}

The global aim here is to assess the emotion recognition performance once a speaker anonymization system has transformed the voices.
Thus, the following sections present the emotional dataset, the evaluation protocol (based on the VPC attack scenarios), and the evaluation results.

\begin{table*}[ht]
    \centering
    \caption{$\mathit{WER}$ and $\mathit{UAR}$ results on IEMOCAP. The LibriSpeech results from VPC are presented for comparison purposes. The first line shows the $\mathit{WER}$ and $\mathit{UAR}$ results when no anonymization is performed. The second line shows the corresponding results when speech is anonymized and evaluated using an ASR system retrained on anonymized speech and an \textit{Informed} attacker scenario.}
    \begin{tabular}{c c c c}
    \toprule
    \multirow{2}{*}{} & \multicolumn{2}{c}{{$WER_\%$}} &
    {$\mathit{UAR_\%}$} \\ 
        {} &
        \centering{LibriSpeech} & \centering{IEMOCAP} & {IEMOCAP} \\
    \midrule
    Original speech data & \multirow{2}{*}{4.15} & \multirow{2}{*}{34.62} & \multirow{2}{*}{44.48} \\
    Model trained on original speech \\
    \midrule
    Anonymized speech data & \multirow{2}{*}{4.77 }& \multirow{2}{*}{38.97} & \multirow{2}{*}{37.92}\\
    Model trained on anonymized speech \\
    \midrule
    Difference Anonymized $/$ Original & 15\% degradation & 13\% degradation & 15\% degradation \\
    \bottomrule
    \end{tabular}
    \label{tab:speechUtility}
\end{table*}

\begin{figure}[h]
\centering
\includegraphics[width=0.90\linewidth]{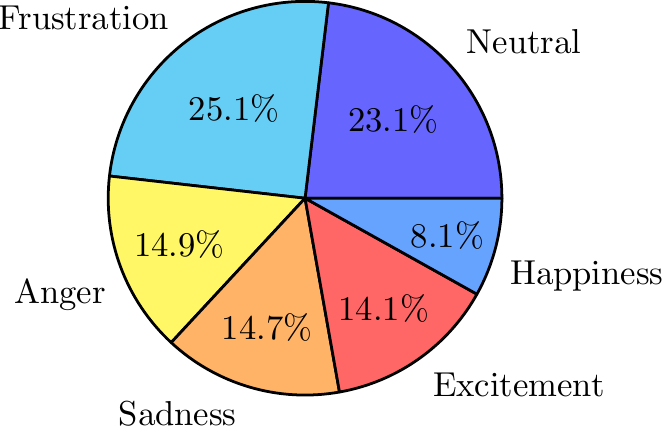}
\caption{Original distribution of emotion categories in the IEMOCAP dataset. Percentages are displayed with respect to number of utterances.}
\label{fig:Emo_balance}
\vspace{-0.6cm}
\end{figure}

\subsection{Dataset}

The IEMOCAP dataset~\cite{Busso08} is an emotional dataset used for experiment purposes such as emotion recognition and speech recognition.
It is composed of 12h of audio-visual data. Improvised and scripted dialogues between 10 female and male actors in the English language were recorded.
Directional microphones have been used to capture each speaker's speech.
It implies that in audio files, the two speakers can appear simultaneously (overlap).
In the case of overlapping speech, the closest speaker to the microphone is considered dominant, and its speech will only be transcribed in the reference transcriptions.
Because of directional microphones used, the level of the overlapping voice is much lower than the level of the "dominant" speaker voice.

The data is segmented by (dominant) speaker turns.
Each turn has been annotated with emotion categories by six human annotators.
Only the recordings that had the majority of annotators agreed on were used.
Figure~\ref{fig:Emo_balance} shows the distribution of emotional labels in the dataset.
Following previous works \cite{Pappagari2020XVectorsME,asr_IEMOCAP}, we consider only five emotions: neutral, frustration, sadness, anger, and happiness.
Happiness combines the original annotations of happiness and excitement to balance the number of utterances in each emotion class.

\subsection{Evaluation protocol}
\label{sub:eval_proto}

In this paper, we focus on evaluating emotional information present in the speech signal, both in the original utterances and in the corresponding anonymized utterances. These evaluations are carried out using a speech emotion recognition system.

The emotion recognition system used is based on a Support Vector Machine (SVM) model. 
In the SER literature, SVM with the radial basis function non linear kernel is widely used \cite{lee2011svmIemocap,Tahon2016} as baseline.
As input features, the eGeMAPS features are used as they provide a minimalist yet efficient representation of emotion \cite{GEMAPS}.
We also experimented MFCCs as input, and the results are similar.
Following the VPC attack scenarios, the emotion recognition is evaluated using the \textit{Ignorant} and the \textit{Informed} attacker scenarios.
In the \textit{Ignorant} scenario, the SVM model used to evaluate the emotion information is trained on non-anonymized original speech data.
For the \textit{Informed} scenario, the SVM is trained on anonymized speech data.
The standard Unweighted Average Recall ({$\mathit{UAR}$}) metric score (defined in Equation~\ref{eq_uar}) is used to measure the emotion recognition performance.
High {$\mathit{UAR}$} values means good emotion recognition.
Regardless of the attack scenario, and to accommodate for the small dataset, the training and evaluation are performed using leave-one-session-out cross-validation protocol.
The global performance is obtained by computing the $\mathit{UAR}$ globally on the five test folds.

\begin{equation}
    \label{eq_uar}
   \mathit{UAR} = \frac{\sum \text{Recall per class}}{\#~ class}
\end{equation}

Utility evaluation, which assesses the preservation and intelligibility of the linguistic content, is performed using two Automatic Speech Recognition (ASR) systems provided by the VPC organizers.  Results are reported with the Word Error Rate ($\mathit{WER}$).
The lower the $\mathit{WER}$ is, the more intelligible the speech is.
Evaluation results are obtained using an ASR system trained on original non-anonymized LibriSpeech \textit{train-clean-360} data and a second one trained on the corresponding anonymized data.
Retraining the ASR system on anonymized speech significantly decreases the $\mathit{WER}$ when decoding anonymized speech data in comparison to the case when the ASR model is trained on the original data.

\begin{figure}[h]
  \centering
  \includegraphics[width=0.90\linewidth, clip=true, trim=1.4cm 1.0cm 2.4cm   3.0cm]{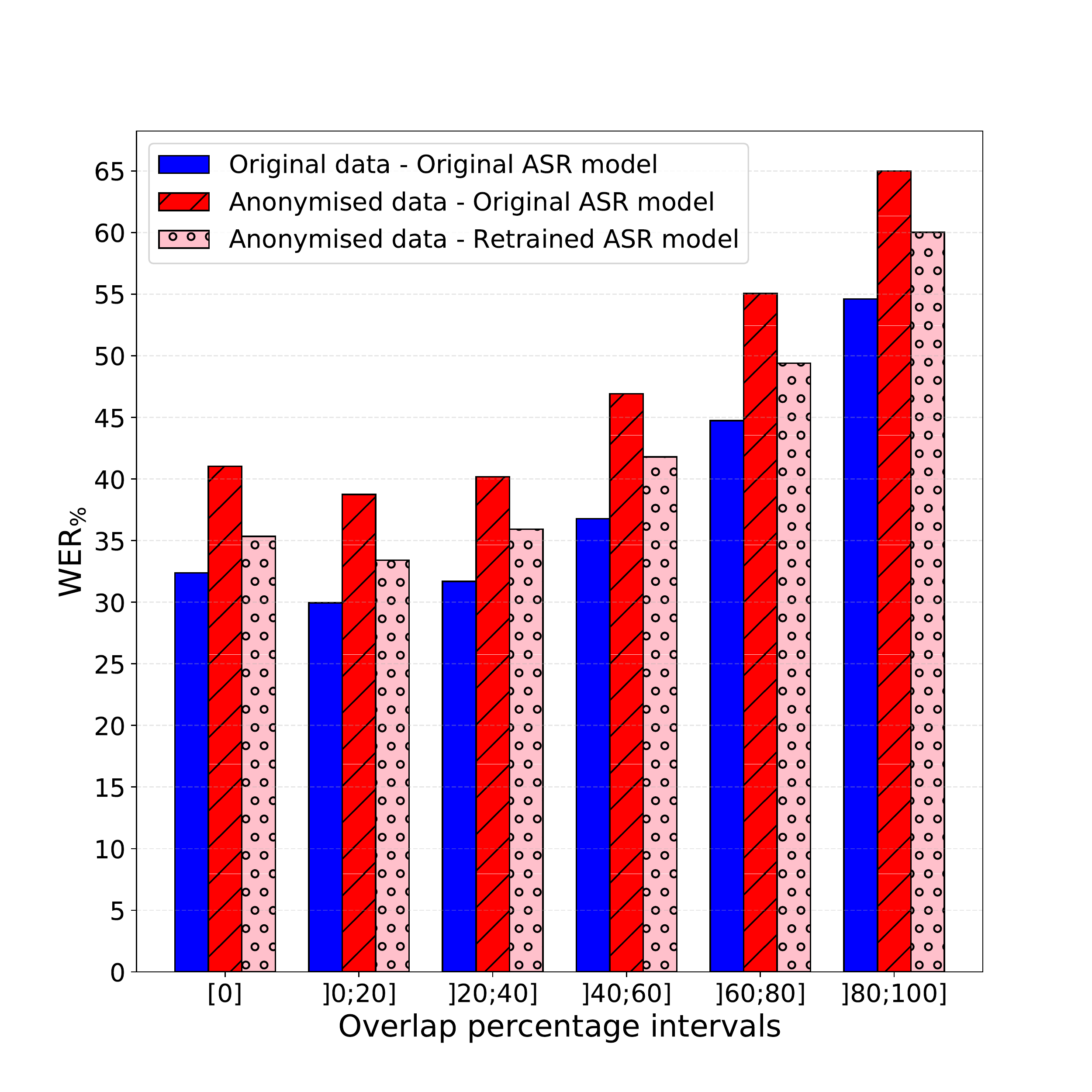}
  \caption{$\mathit{WER}$ performance on IEMOCAP with respect to the amount of overlap speech.
  }
  \label{fig:IEMOCAP_WER.png}
  \vspace{-0.6cm}
\end{figure}

\begin{table*}[ht]
    \caption{Speech emotion recognition ($\mathit{UAR}_\%$) performance for the considered attack scenarios and the 95\% confidence interval.} 
    \centering
    \begin{tabular}{c c c c c c}
    \toprule
         \multicolumn{2}{c}{} & F0 linear          & F0 random   & SVM & \multirow{2}{*}{$\mathit{UAR}_\%$} \\
         \multicolumn{2}{c}{} & transformation  & warping & model          &  \\
    \midrule
         1 & Baseline (original speech)      &                &           & original  & ${44.48 \pm 1.14}$         \\
         \midrule
         2 & \multirow{3}{*}{\begin{minipage}{0.5in}\centering\textit{Ignorant} attacker\end{minipage}}   &                &           & original  & ${21.97 \pm 0.95}$    \\
        3 &    &  \checkmark      &           & original  & ${21.80 \pm 0.94}$   \\
        4 &    &  \checkmark      & \checkmark          & original  & ${22.14 \pm 0.95}$   \\
        \midrule
        5 & \multirow{3}{*}{\begin{minipage}{0.5in}\centering\textit{Informed} attacker\end{minipage}}   &                &           & retrained  & ${37.92 \pm 1.11}$   \\
        6 &    &  \checkmark      &           & retrained  & ${37.88 \pm 1.11}$   \\
        7 &    &  \checkmark      & \checkmark          & retrained  & ${38.84 \pm 1.11}$   \\

    \bottomrule
    \end{tabular}
    \label{tab:results}
\end{table*}

\subsection{Results}
This section presents the experimental results for the evaluation protocol detailed in Section \ref{sub:eval_proto}.

{Table {\ref{tab:speechUtility}}} shows $\mathit{WER}$ results on the IEMOCAP dataset, both for the original speech data, and for the anonymized speech data. Original speech is evaluated using the ASR model trained on original speech, while anonymized speech is evaluated with the model retrained on anonymized speech.
The corresponding LibriSpeech scores from the VPC post-evaluation analysis \cite{VPCPostEval20} are presented for comparison purposes.
In the IEMOCAP dataset, from original to anonymized, the utility score drops from 34.62 {$\mathit{WER}_\%$} to 38.97 {$\mathit{WER}_\%$}, which represents a relative degradation of 13\%.
Similar behavior is observed on the LibriSpeech dataset, meaning the VPC baseline anonymization system performs the transformation properly on emotional speech.
The high $\mathit{WER}$ on IEMOCAP can be explained by the presence of overlaps, and non-neutral speech, which is not present in the LibriSpeech data.

Following the attack scenarios defined by the VPC (see Section \ref{subsec:VPC_scenars}), the emotion recognition scores under the \textit{Ignorant} and \textit{Informed} attackers are summarized in {Table {\ref{tab:results}}}.
For an \textit{Ignorant} attacker, the $\mathit{UAR}$ is 21.97\%, which is nearly equal to random guessing.
For an \textit{Informed} attacker, the $\mathit{UAR}$ is 37.92\%, which is a 15\% degradation compared to the UAR measured on original speech.
The emotion recognition performance in terms of $\mathit{UAR}$ seems to be impacted in the same manner as the utility performance (measured by $\mathit{WER}$).


\subsection{Prosodic parameters}

If one wants to hide emotional information in the anonymized speech, modification of the prosodic parameters (i.e., fundamental frequency, intensity, and rhythm) are needed.
Note that F0 mean, variability, and range are included in the eGeMAPS~\cite{GEMAPS} input features used in our experimental setup.
As F0 was available in the anonymization baseline system, we carried out some experiments involving random modifications of the F0 values (as described in Section \ref{subsubsec:f0_random}).

Results are shown in {Table {\ref{tab:results}}} together with the 95\% confidence interval. 
We can see that, regardless of the attacker scenario, applying or not the F0 linear transformation alone, or the F0 linear transformation followed by a F0 random warping, leads to very similar results in terms of emotion recognition scores ($\mathit{UAR}$).
In the \textit{Informed} attacker scenario, where the SVM classifier is trained on anonymized speech, applying the F0 linear transformation does not modify the emotion recognition performance.
In the future, we will investigate other modifications of the F0 curves in the anonymization process.

\section{Conclusion}

In this paper, we have evaluated the application of the Voice Privacy baseline system on emotional speech.
Concerning the utility metric, based on automatic speech recognition performance, we observed a 13\% degradation of the Word Error Rate ($\mathit{WER}$) on the IEMOCAP anonymized data compared to the $\mathit{WER}$ measured on original speech. This degradation is similar to the one reported on the LibriSpeech data.
However, the $\mathit{WER}$ is much higher on the emotional data (IEMOCAP), also impacted by the presence of overlapping speech, than on the clean neutral data (LibriSpeech).

For what concerns the emotion information carried by the speech signal, we have measured it through the standard Unweighted Average Recall ($\mathit{UAR}$) metric. We have observed a 15\% degradation of the $\mathit{UAR}$ when measured on the anonymized data compared to its measure on original data.
The degradation observed for emotion recognition is similar to the degradation observed on the Word Error Rate (that measures the utility), which is fine if one considers the emotion a valuable information to be kept in the anonymized speech signal.

However, one can also consider emotion as personal information that the anonymization system should remove. The preliminary experiments reported in this paper regarding simple random modifications of the F0 values show that such simple modifications are not enough to hide emotional information. Hence, further research will investigate other modifications of the F0 values and modifications of the duration and energy, which are other prosodic parameters that carry emotion information.

\section{Acknowledgements}
This work was made with the support of the French National Research  Agency, in the framework of the project ANR DEEP-PRIVACY (18-CE23-0018) and Région Grand Est.
Experiments presented in this paper were carried out using the Grid'5000 testbed, supported by a scientific interest group hosted by Inria and including CNRS, RENATER and several Universities as well as other organizations (see https://www.grid5000.fr).

\bibliographystyle{IEEEtran}

\bibliography{main}

\end{document}